\documentclass[a4paper,twocolumn]{esapub} 
\usepackage{epsfig} 		
\usepackage{natbib}

\def \sig      {$\sigma$}
\def \gray     {$\gamma$-ray}
\def \grays    {$\gamma$-rays}

\title{
Multifrequency Observations of the Gamma-ray Blazar 3C 279 in Low-State
during INTEGRAL AO-1
}
\author[1]{W.~Collmar}
\author[2]{M.~B\"ottcher}
\author[1]{V.~Burwitz}
\author[3]{T.~Courvoisier}
\author[1]{S.~Komossa}
\author[1]{P.~Kretschmar}
\author[5]{E.~Nieppola}
\author[4]{K.~Nilsson}
\author[5]{T.~Ojala}
\author[1]{K.~Pottschmidt}
\author[4]{M.~Pasanen}
\author[4]{T.~Pursimo}
\author[4]{A.~Sillanp\"a\"a}
\author[4]{L.~Takalo}
\author[5]{M.~Tornikoski}
\author[6]{H.~Ungerechts}
\author[7]{E.~Valtaoja}
\author[3]{R.~Walter}
\author[9]{R.~Webster}
\author[8]{M.~Whiting}
\author[7,10]{K.~Wiik}
\author[9]{I.~Wong}

\affil[1]{Max-Planck-Institut f\"ur extraterrestrische Physik,P.O. Box 1312,  85741 Garching, Germany}
\affil[2]{Department of Physics and Astronomy, Ohio University, Athens, OH 45701, USA }
\affil[3]{INTEGRAL Science Data Center, Chemin d'E'cogia 16, 1290 Versoix, Switzerland}
\affil[4]{Tuorla Observatory, V\"ais\"al\"antie 20, FI-21500 Piikki\"o, Finland}
\affil[5]{Mets\"ahovi Radio Observatory, Helsinki University of Technology, 02540 Kylm\"al\"a, Finland}
\affil[6]{IRAM, Avenida Divina Pastora 7, Nucleo Central, E-18012 Granada, Spain}
\affil[7]{Department of Physical Sciences, University of Turku, 20100 Turku, Finland}
\affil[8]{Department of Astrophysics and Optics, University of New South Wales, Sydney NSW 2052, Australia}
\affil[9]{Astrophysics Group, University of Melbourne, Victoria 3010, Australia}
\affil[10]{ISAS, VSOP-Group, 3-1-1 Yoshinodai, Sagamihara, Kanagawa 229-8510, Japan}

\begin{document}


\maketitle

\begin{abstract}
We report first results of a multifrequency campaign from radio to hard X-ray 
energies of the prominent \gray\ blazar 3C~279 during the first year of the INTEGRAL 
mission. The variable blazar was found at a low activity level, but 
was detected by all participating instruments. Subsequently a multifrequency 
spectrum could be compiled. The individual measurements as well as 
the compiled multifrequency spectrum are presented. In addition, this 
2003 broadband spectrum is compared to one measured in 1999 during a 
high activity period of 3C~279.  
\end{abstract}

\section{Introduction}

The EGRET experiment aboard CGRO has identified about 90
blazar-type AGN emitting high-energy ($>$100~MeV) $\gamma$-rays
\citep{Hartman99}. Among the most prominent ones is 3C~279, 
an optically violently variable (OVV) quasar, located at a redshift of 0.538.
It shows rapid variability in all
wavelength bands, polarized emission in radio and optical,
superluminal motion, and a compact radio core with a flat radio spectrum. 
These properties put the quasar 3C~279 into the blazar sub-class 
of AGN. According to the unified model of AGN, blazars are sources which 
expel jets close to our line-of-sight. The blazar emission is 
predominantly of non-thermal origin, showing a typical two-hump spectrum 
from radio to \grays. It is generally believed that in blazars the 
radio--through-optical/UV continuum is synchrotron radiation generated by
relativistic electrons in a magnetized jet and the high-energy
 continuum from 
X-rays to \grays\ is due to Comptonization of soft photons
 by the non-thermal jet electrons.
Most emission models have been developed within this scenario.
Key predictions of the different models which can be tested by multifrequency 
observations are the predicted spectral shapes for blazars from radio to \gray\ energies,
and the predicted temporal variations as function of energy.  
  
3C~279 was detected by EGRET several times \citep[e.g.,][]{Hartman01}, showing 
strong flaring activity with flux changes up to a factor of 100. 
3C~279 was also detected by CGRO/COMPTEL at low-energy (1-30~MeV) \grays\ 
\citep[e.g.,][]{Williams95} and CGRO/OSSE (50 keV - 1 MeV) at the transition 
from hard X-rays to \grays\ \citep{NaronBrown95}, i.e.,
throughout the whole INTEGRAL band and above. 
Because of the low-significance OSSE detections, which only occurred on or near
flaring periods \citep{NaronBrown95}, not much is known on the 
hard X-ray properties of 3C~279. To improve on that INTEGRAL observed 
the blazar in its AO-1 period. To correlate the hard X-ray and multifrequency 
properties of 3C~279, the INTEGRAL observations were supplemented
by multifrequency coverage, i.e., contemporaneous radio and mm, near-infrared and
 optical observations, as well as X-ray observations.    

In this paper we present first results of our multiwavelength campaign on 3C~279 
in 2003 (not all data are available yet). Because of the limited page number,
we concentrate on presenting the main observational results. 
A more detailed presentation, including a discussion on the scientific implications
of the new results, will be given in a later paper (Collmar et al., in prep.).

\section{OBSERVATIONS}

3C~279 was observed continuously for 300 ksec in INTEGRAL AO-1 between
June 1 and June 5, 2003. These high-energy observations were supplemented 
in X-rays by a short Chandra pointing of 5 ksec on June 2, and by ground based
monitoring from radio to optical bands, including the following measurements: 
37~GHz at the Mets\"ahovi radio telescope, 250~GHz at the SEST, 3 mm, 2 mm, and 1.3 mm 
at the 30m IRAM telescope, optical R-band monitoring at the 60cm Tuorla telescope 
at La Palma, and UBVRI photometry at the 2.3m telescope at Siding Spring, Australia. 
In addition, VLBA observations at six radio frequencies from 5 to 86 GHz 
have been carried out and the resulting data is currently being reduced. 
 
\begin{figure}[th]
\centering
\epsfig{figure=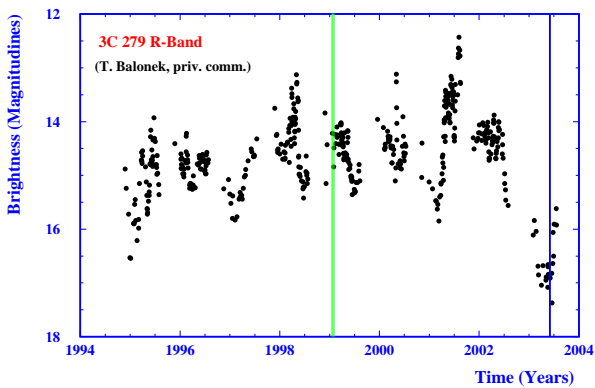,width=8.0cm,clip=}
\caption{
Longterm R-band light curve of 3C~279 during the last ten years.
The right line indicates the observation time of 
INTEGRAL in 2003, and the left line corresponds to a \gray\ high state 
observation in 1999 (see section 3.3). The INTEGRAL observation was carried
out during the lowest R-band flux state of the last 10 years.
The figure is from \citet{Balonek04}. 
\label{fig:1}
}
\end{figure}

The data analysis revealed that 3C~279 was observed in a deep low-activity
state. This is obvious from Fig.~\ref{fig:1}, which shows the longterm optical R-band
light curve of the blazar during the last 10 years. According to CGRO experience, 
the optical R-band is a good tracer for the high-energy activity of blazars.
The campaign was carried out during the faintest R-band brightness
of the last 10 years, roughly 5 mag fainter than the maximum, 
and about 2.5 to 3 mag fainter than average.   

\section{RESULTS}
\subsection{High-Energy Observations}
\begin{figure}[th]
\centering
\epsfig{figure=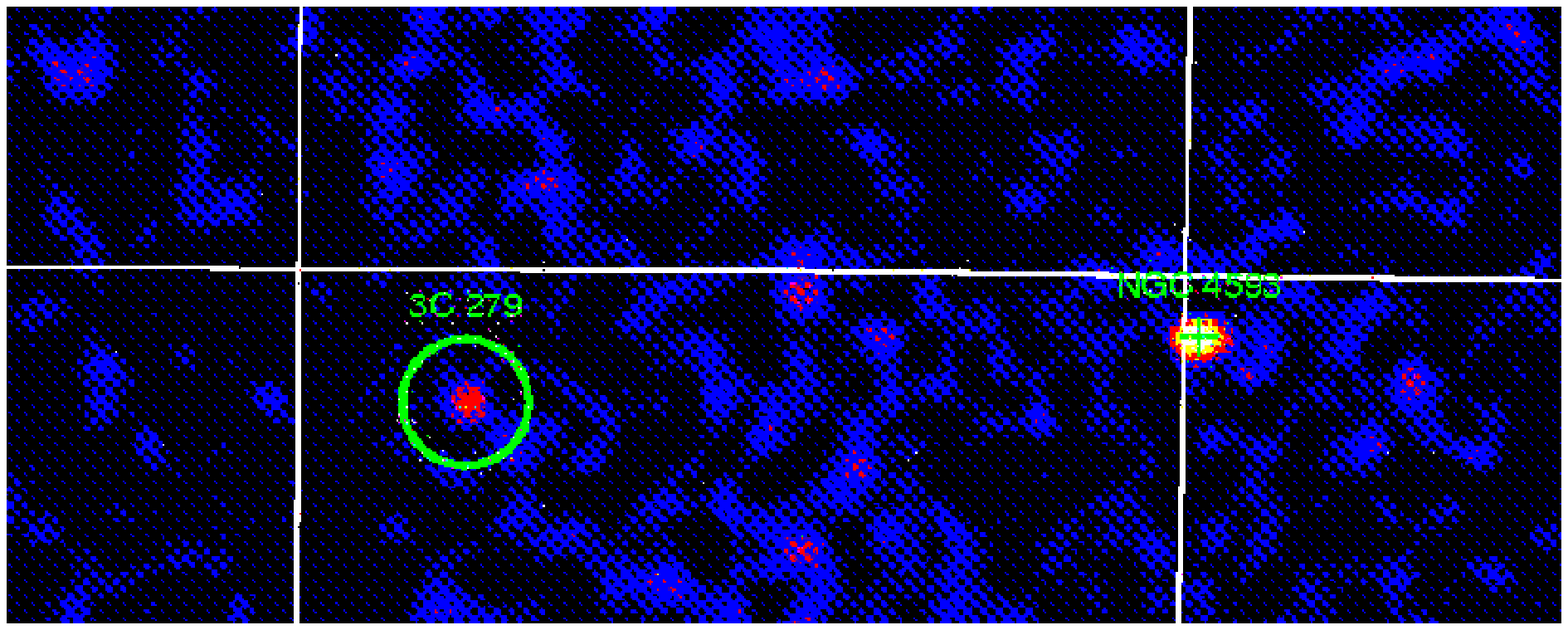,width=8.0cm,clip=}
\epsfig{figure=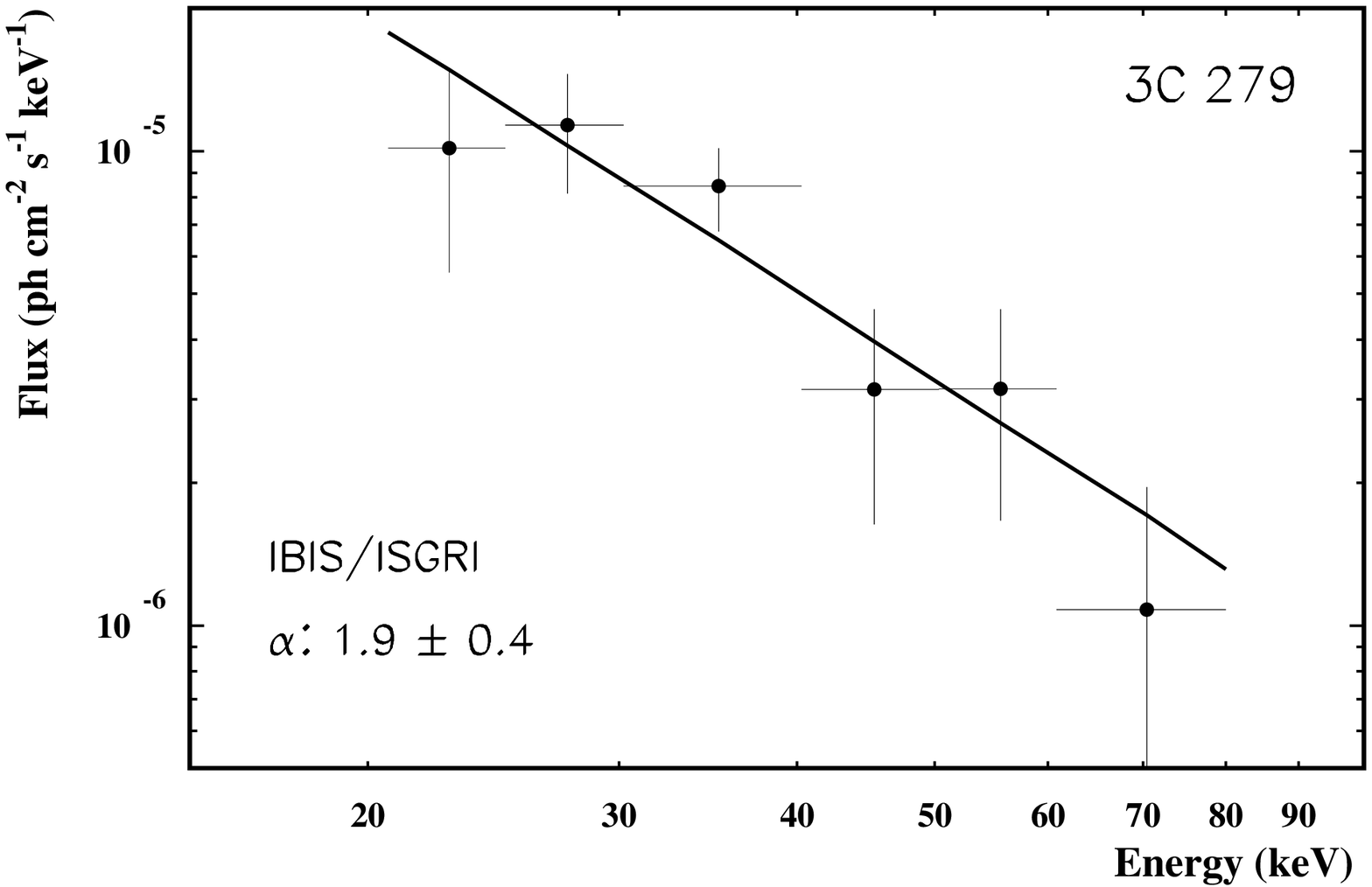,width=8.0cm,clip=}
\caption{
Top: The ISGRI image shows a 5.4\sig\ detection of 3C~279 in its 
sensitive 30-50~keV band. In addition, the Seyfert galaxy 
NGC 4593 is even more clearly detected.
\newline Bottom: The ISGRI hard X-ray spectrum between 20 and 80 keV
is shown, together with the best-fit power-law shape.  
\label{fig:2}
}
\end{figure}

The IBIS/ISGRI experiment aboard INTEGRAL detected the \gray\
blazar at energies between 20 and 80~keV with a formal significance of 6.6\sig.
Above 80~keV no signs of 3C~279 could be found yet in the IBIS data.
The ISGRI spectrum between 20 and 80~keV, 
compiled in 6 spectral bins, is, by assuming a power-law shape, consistent with
a photon index of 1.9$\pm$0.4 (1\sig). INTEGRAL/SPI and 
-JEM-X analyses did not yet reveal a significant detection
of the source, although marginal evidence is found in the JEM-X data.
The ISGRI detection is, together with the CGRO/OSSE 
detection in 1991 (about 10\sig\ in one 50-150~keV bin; McNaron-Brown et al. 1995) 
the only significant detection at hard X-rays. 
In particular, the spectral shape of 3C~279 could be measured in the
 20 to $\sim$100~keV
range for the first time, although the blazar was in low-activity state.
Given this result, a hard X-ray spectrum of unprecedented quality for 3C~279 
should be measured in a high-state observation for a similar 
observation time.  
The ISGRI image and spectrum are shown in Fig.~\ref{fig:2}.

\begin{figure}[th]
\centering
\epsfig{figure=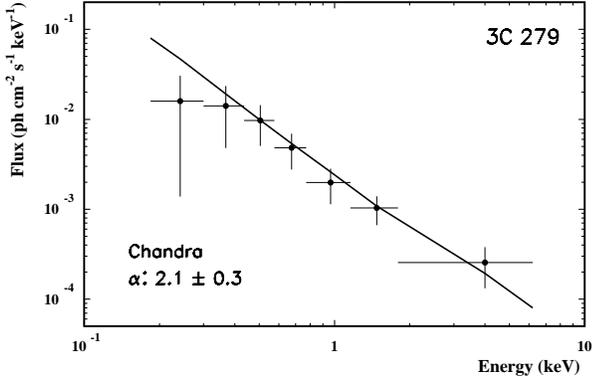,width=8.0cm,clip=}
\caption{
The Chandra 0.2 - 6 keV X-ray spectrum is shown,
together with the best-fit power-law shape.
\label{fig:3}
}
\end{figure}

The contemporaneous short Chandra pointing was carried out with the 
LETGS (Low Energy Transmission Grating Spectrograph). This setup was used
to search for a possible soft excess. In fact, the goal was 
to measure the crossover point of the anticipated {\it low-energy} 
synchrotron and {\it high-energy} inverse-Compton emission
by using the improved soft X-ray sensitivity of Chandra compared to 
previous X-ray missions. 
The energy of this crossover point is inferred to be at 
soft X-ray energies by modeling the 3C~279 broadband emission 
\citep [e.g.,][]{Hartman01}, but has never been observed yet. 
The blazar is clearly visible in the 0th order of the LETGS, 
however, is weak in its 1st order, which is necessary to generate an 
energy spectrum. Therefore only a weakly determined spectrum 
between 0.2 and 6~keV could be derived. Assuming the canonical 
power-law shape at X-ray energies, spectral fitting yields a best-fit
shape (Fig.~\ref{fig:3}) of photon index 2.1$\pm$0.3 (1\sig). 
This value suggests a softer spectrum than usually measured in X-rays, 
however, is still consistent with the canonical X-ray power-law
slope of photon index of $\sim$1.7. In particular, no soft excess
is found in the data.         

\begin{figure}[th]
\centering
\epsfig{figure=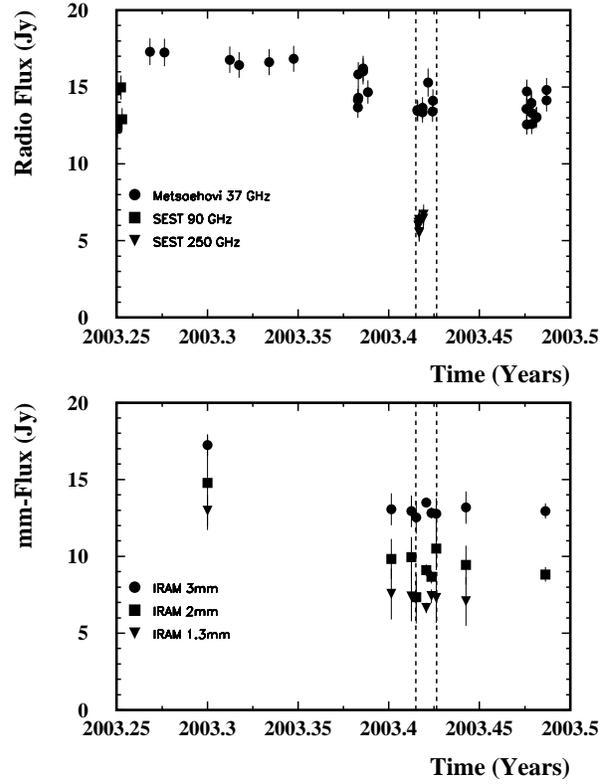,width=8.0cm,clip=}
\caption{
Radio and mm light curves from monitoring observations 
at or around the INTEGRAL observational period, which is
indicated by the vertical dashed lines.
\label{fig:4}
}
\end{figure}

\begin{figure}[th]
\centering
\epsfig{figure=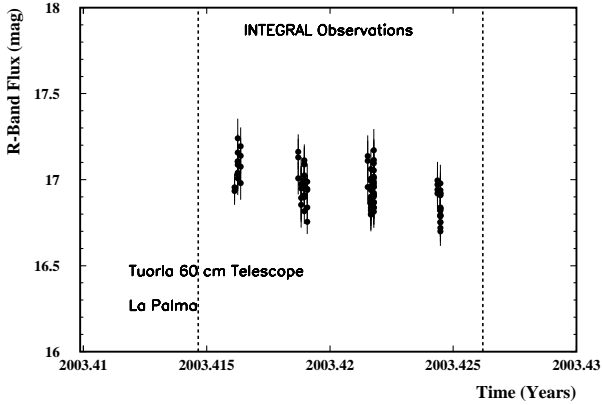,width=8.0cm,clip=}
\caption{Daily R-band monitoring during the campaign
by the Tuorla 60cm telescope. Noteable is 
the surprisingly steady flux during the 4-day period. 
The INTEGRAL observational period is 
indicated by the vertical dashed lines.   
\label{fig:5}
}
\end{figure}

\subsection{Low-Energy Observations}
Because 3C~279 is a prominent blazar, it is monitored regularly in radio,
IR and optical bands, probably at least on a monthly basis or so.
However, during our campaign the monitoring was denser, and, in particular,
was carried out contemporaneously
in the different bands including large telsecopes (e.g., 2.3m Siding Spring telescope, 
30m IRAM telescope) to derive accurate flux estimates. Some monitoring results
are shown in Figs.~\ref{fig:4} and \ref{fig:5}. Generally, the results of
the ground-based monitoring can be summarized as follows: 
1) the flux of 3C~279 was surprisingly stable, 
and 2) the flux level was generally lower than average,
especially in the optical bands.

\subsection{Multiwavelength Results}
\begin{figure}[th]
\centering
\vspace*{-0.5cm}
\epsfig{figure=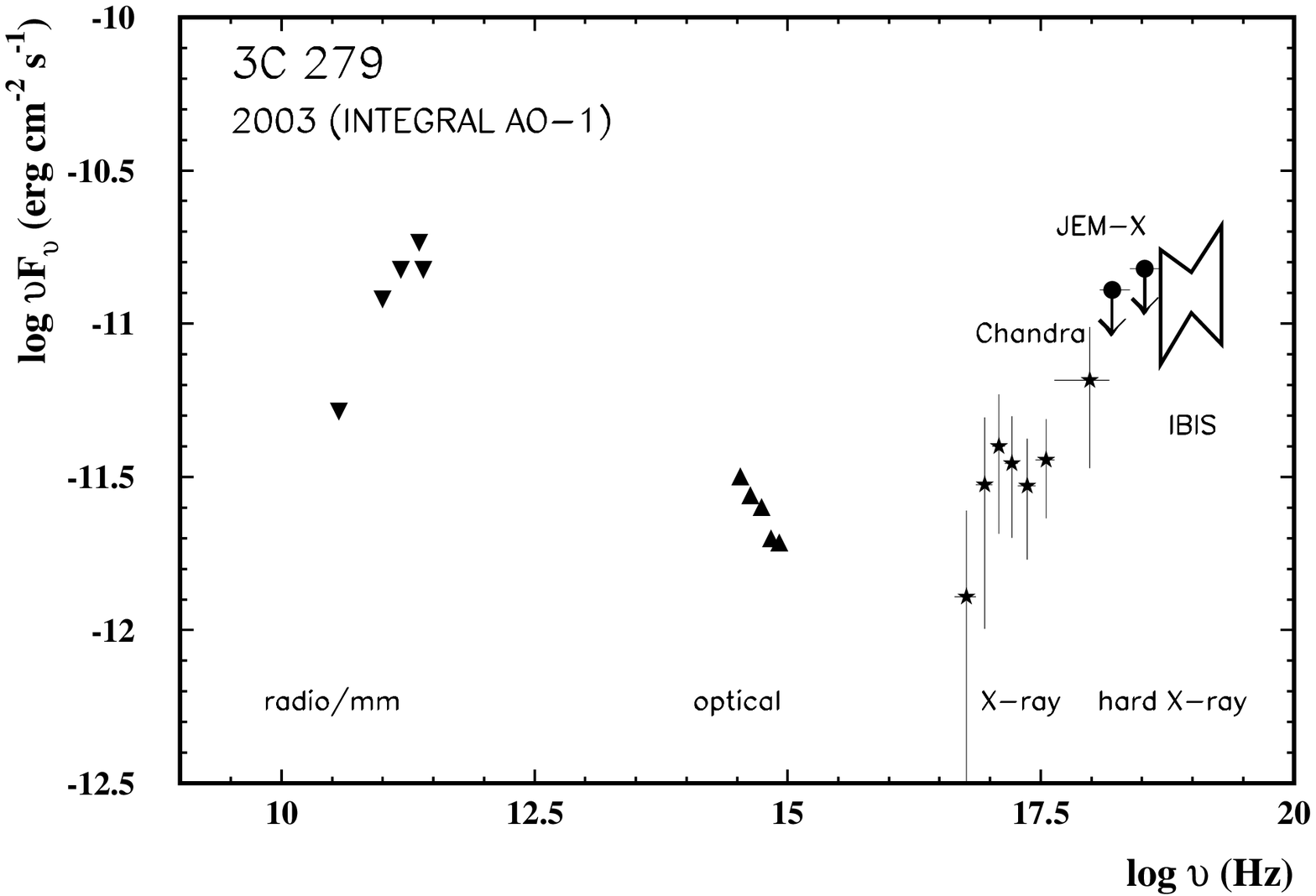,width=8.0cm,clip=}
\epsfig{figure=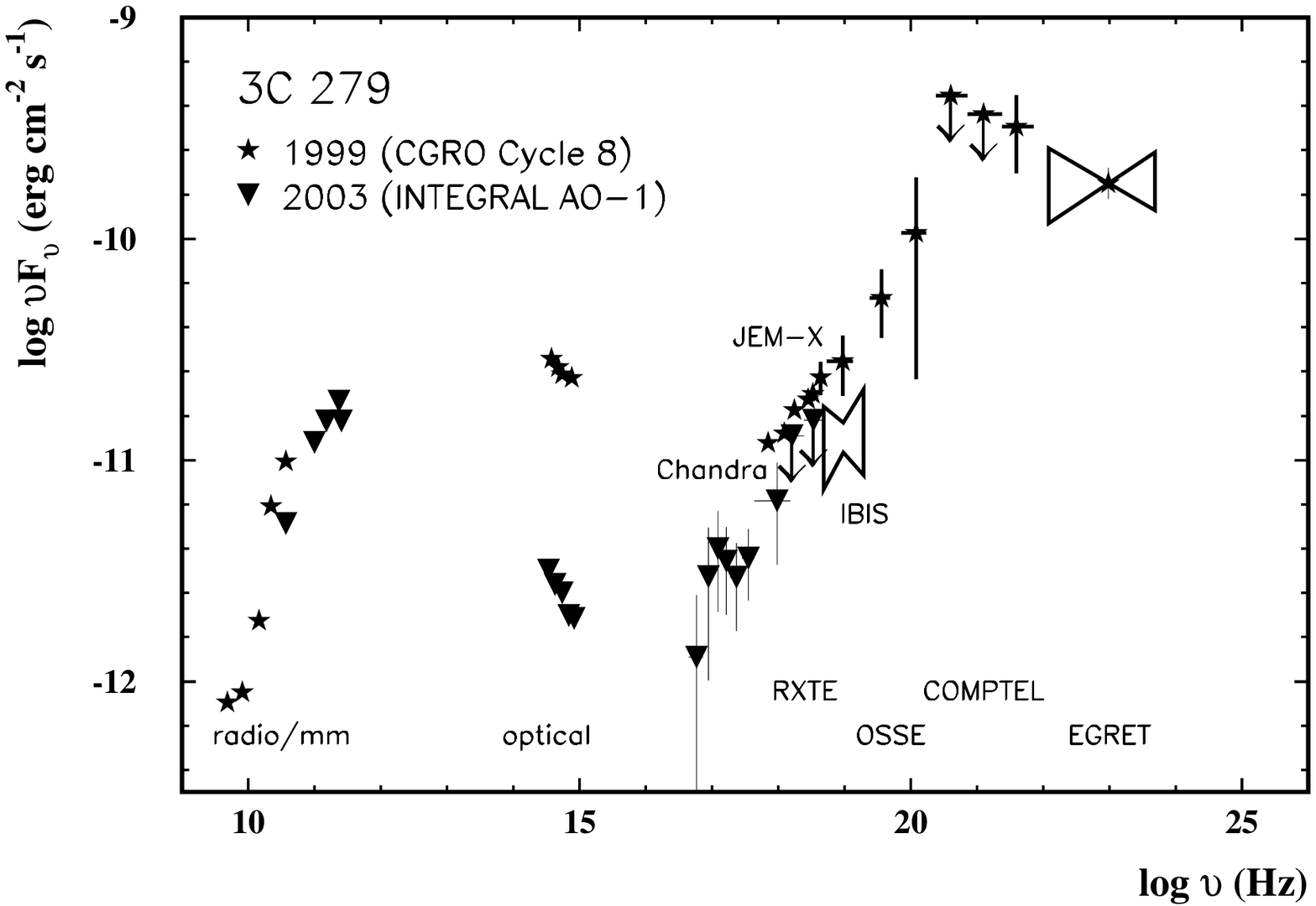,width=8.0cm,clip=}
\caption{
Top: The 3C~279 multifrequency spectrum as measured during the campaign in 
June 2003. The spectrum shows the typical two-hump shape, with a crossover
between the optical U band and 0.2~keV. Flux points are shown with 1\sig\ error bars. 
Upper limits are 2\sig. For ISGRI the error contours on the spectral 
shape are given. The JEM-X upper limits correspond to the bands 5-10 keV and 
10-20~keV. At low energies the error bars are smaller
than the symbols. 
\newline Bottom: Comparison of the 2003 low-state spectrum to 
the high-state multiferquency spectrum of 1999 (Collmar et al. 2000).
In particular,  the optical fluxes are more than an order of magnitude
different.
\label{fig:6}
}
\end{figure}

Because the different measurements were collected contemporaneously
(i.e., within the INTEGRAL observational period of 4 days), 
a multifrequency spectrum from the radio to the hard X-ray band could be compiled
(Fig.~\ref{fig:6}, top). In case of serveral measurements during the INTEGRAL period
(e.g., the IRAM mm measurements), the fluxes were averaged. This spectrum 
shows the typical two-hump shape, which is believed to be synchrotron 
at the lower - and inverse-Compton (IC) emission at the higher energies. 
The radio and mm fluxes are, as usual, on the rising branch of the 
synchrotron emission, whose actual maximum, probably located at IR energies, 
is not observed. The optical data points (UBVRI photometry at Siding Spring)
are on the decaying branch of the synchrotron emission, showing
almost a perfect power-law
shape. The X-ray and hard X-ray fluxes are already on the rising IC branch of the 
emission. Because Chandra did not find a soft excess, the crossover point of 
the synchrotron and IC emission is not observed. Therefore it has to be 
lower than 0.2~keV. 
This low-state multifrequency spectrum generally resembles the one
measured by \citet{Maraschi94} during another low-state period of 
3C~279 in December/January 1992/93, still including the CGRO experiments.
However, by the participation of Chandra and INTEGRAL,
the 2003 one is covered much better at X- and hard X-ray energies.
A comparison of these 2 spectra will be scientifically interesting; the two 
spectra might even complement each other.

In Fig.~\ref{fig:6} (bottom), the measured low-state is compared to a high-state
measurement during the CGRO era in 1999 \citep{Collmar00}. 
At radio and mm bands the fluxes are at the same 
level, though if comparable, lower than during the high-state observation. 
The main deficit occurs in the optical bands, where the flux is more
than 1 order of magnitude below the high-state observation.
At soft X-ray energies, not covered during the high state, the Chandra measurements 
indicate a deep minimum, located somewhere in the UV.
Above 2 keV (covered by RXTE during the high state), the flux level approaches 
the high state one. The new and surprising fact is, 
that despite the large differences in 
optical flux, the hard X-ray flux, as measured by ISGRI, is quite close to 
the high-state measurement. The main deficit occurs from at least
the optical to the X-ray band. Something happens which suppresses
the synchrotron flux at optical energies but only weakly affects
the SSC flux at hard X-rays. This result provides at least new constraints 
for the modelling of 3C~279.      

\section{Summary}
A multiwavelength campaign on the prominent and variable 
\gray\ blazar 3C~279, including INTEGRAL, was carried out in 2003.
According to the longterm optical R-band lightcurve, 
the source was observed in its lowest optical activity state of the last decade.
Because the source was detected by all participating instruments, 
a scientifically interesting low-state multifrequency spectrum could be 
compiled from radio to hard X-ray energies. It has an unprecedented coverage
in the high-energy domain by Chandra and INTEGRAL, and provides new constraints 
for the modelling of 3C~279.

\end{document}